\documentclass[12pt]{article}
\usepackage[numbers,square,sort]{natbib}
\usepackage[utf8]{inputenc}
\usepackage{amsmath}
\usepackage{graphicx}
\usepackage{amsfonts}
\usepackage[left=2cm,top=2cm,right=2cm]{geometry}
\usepackage{setspace}
\usepackage{verbatim}
\newcommand{\otherpaper}{\citep{3989Wittmann2013}}

\title{Population genetic consequences of the Allee effect and the role of offspring-number variation}
\author{Meike J. Wittmann, Wilfried Gabriel, Dirk Metzler}
\date{}

\begin{document}
\maketitle

\begin{abstract}
\noindent A strong demographic Allee effect in which the expected population growth rate is negative below a certain critical population size can cause high extinction probabilities in small introduced populations. However, many species are repeatedly introduced to the same location and eventually one population may overcome the Allee effect by chance. With the help of stochastic models, we investigate how much genetic diversity such successful populations harbour on average and how this depends on offspring-number variation, an important source of stochastic variability in population size. We find that with increasing variability, the Allee effect increasingly promotes genetic diversity in successful populations. Successful Allee-effect populations with highly variable population dynamics escape rapidly from the region of small population sizes and do not linger around the critical population size. Therefore, they are exposed to relatively little genetic drift. We show that here---unlike in classical population genetics models---the role of offspring-number variation cannot be accounted for by an effective-population-size correction. Thus, our results highlight the importance of detailed biological knowledge, in this case on the probability distribution of family sizes, when predicting the evolutionary potential of newly founded populations or when using genetic data to reconstruct their demographic history.
\end{abstract}

\vspace{0.2cm}


\pagebreak

\section{Introduction}
The demographic Allee effect, a reduction in per-capita population growth rate at small population sizes \citep{2080Stephens1999}, is of key importance for the fate of both endangered and newly introduced populations, and has inspired an immense amount of empirical and theoretical research in ecology \citep{3987Courchamp2008}. By shaping the population dynamics of small populations, the Allee effect should also strongly influence the strength of genetic drift they are exposed to and hence their levels of genetic diversity and evolutionary potential. In contrast to the well-established ecological research on the Allee effect, however, research on its population genetic and evolutionary consequences is only just beginning \citep{3699Kramer2008,605Hallatschek2008,3635Roques2012}. In this study, we focus on the case where the average population growth rate is negative below a certain critical population size. This phenomenon is called a strong demographic Allee effect \citep{2084Taylor2005}. Our goal is to quantify levels of genetic diversity in introduced populations that have successfully overcome such a strong demographic Allee effect. Of course, the population genetic consequences of the Allee effect could depend on a variety of factors, some of which we investigated in \otherpaper. Here we focus on the role of variation in the number of offspring produced by individuals or pairs in the population. 

There are several reasons why we hypothesise offspring-number variation to play an important role in shaping the population genetic consequences of the Allee effect. First, variation in individual offspring number can contribute to variability in the population dynamics and this variability influences whether and how introduced populations can overcome the Allee effect. In a deterministic model without any variation, for instance, populations smaller than the critical size would always go extinct. With an increasing amount of stochastic variability, it becomes increasingly likely that a population below the critical population size establishes \citep{301Dennis2002}. Depending on the amount of variability, this may happen either quickly as a result of a single large fluctuation or step-by-step through many generations of small deviations from the average population dynamics. Of course, the resulting population-size trajectories will differ in the associated strength of genetic drift. Apart from this indirect influence on genetic diversity, offspring-number variation also directly influences the strength of genetic drift for any given population-size trajectory. In offspring-number distributions with large variance, genetic drift tends to be strong because the individuals in the offspring generation are distributed rather unequally among the individuals in the parent generation. In distributions with small variance, on the other hand, genetic drift is weaker. 

In  \otherpaper, we have studied several aspects of the population genetic consequences of the Allee effect for Poisson-distributed offspring numbers, a standard assumption in population genetics. However, deviations from the Poisson distribution have been detected in the distributions of lifetime reproductive success in many natural populations. Distributions can be skewed and multimodal \citep{3977Kendall2010} and, unlike in the Poisson distribution, the variance in the number of surviving offspring is often considerable larger than the mean, as has been shown for example for tigers \citep{3972Smith1991}, cheetahs \citep{3980Kelly1998}, and steelhead trout \citep{3959Araki2007}. Several sources contribute to this variance, for example variation in environmental conditions, sexual selection, and predation. For several bird species, there is evidence that pairs individually optimise their clutch size given their own body condition and the quality of their territories \citep{3965Hogstedt1980,3979Davies2012}. Variation in offspring number may also depend on population size or density and thus interact with an Allee effect in complex ways. A mate finding Allee effect, for example, is expected to lead to a large variance in reproductive success among individuals \citep{302Kramer2009} because many individuals do not find a mating partner and thus do not reproduce at all, whereas those that do find a partner can take advantage of abundant resources and produce a large number of offspring. 
In this study, we therefore investigate how the genetic consequences of the Allee effect depend on offspring-number variation. With the help of stochastic simulation models, we generate population-size trajectories and genealogies for populations with and without Allee effect and with various offspring-number distributions, both models with a smaller and models with a larger variance than the Poisson model. 

Although probably only few natural populations conform to standard population genetic assumptions such as that of a constant population size and a Poisson-distributed number of offspring per individual, many populations still behave as an idealised population with respect to patterns of genetic variation \citep{619Charlesworth2009}. The size of this corresponding idealised population is called the effective population size and is often much smaller (but can, at least in theory, also be larger) than the size of the original population, depending on parameters such as the distribution of offspring number, sex ratio, and population structure. Because of this robustness and the tractability of the standard population genetics models, it is common to work with these models and effective population sizes, instead of using census population sizes in conjunction with more complex and realistic models. For example, when studying the demographic history of a population, one might estimate the effective current population size, the effective founder population size etc. If one is interested in census population size, one can then use the biological knowledge to come up with a conversion factor between the two population sizes. Therefore, if we find differences in the genetic consequences of the Allee effect between different family size distributions but it is possible to resolve these differences by rescaling population size or other parameters, those differences might not matter much in practice. If, on the other hand, such a simple scaling relationship does not exist, the observed phenomena would be more substantial and important in practice.
We therefore investigate how closely we can approximate the results under the various offspring-number models by rescaled Poisson models.

\section{Methods}
\subsection{Scenario and average population dynamics}
\begin{figure}
  \centering
  \includegraphics{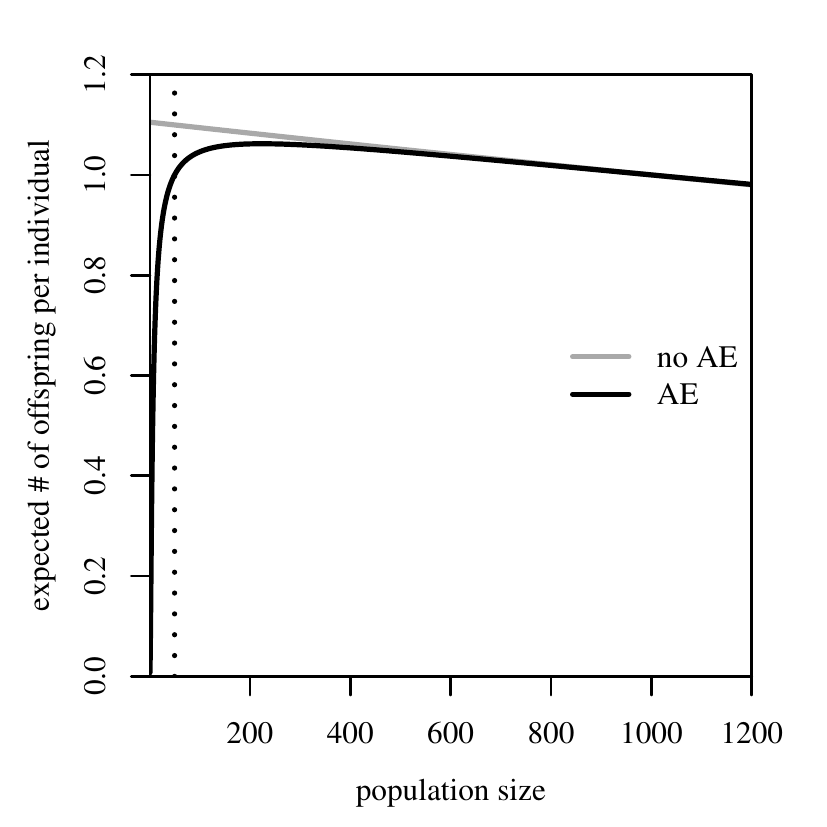}
  \caption{The expected number of surviving offspring per individual (see equation \eqref{eq:expectedpopsize}) as a function of the current population size without Allee effect (no AE, grey line) or with an Allee effect (AE, black line) and critical size $a=50$ (indicated by a dotted vertical line). $k_1=1000, r=0.1$.}
  \label{fig:Alleemodel}
\end{figure}

In our scenario of interest, $N_0$ individuals from a large source population of constant size $k_0$ migrate or are transported to a new location. The average population dynamics of the newly founded population are described by a modified version of the Ricker model \citep[see e.g.][]{1041Kot2001} with growth parameter $r$, carrying capacity $k_1$, and critical population size $a$. Given the population size at time $t$, $N_t$, the expected population size in generation $t+1$ is
\begin{equation}
\mathbf{E}[N_{t+1}|N_t] = N_t \cdot \exp\! \left\{r\cdot \left(1-\frac{N_t}{k_1}\right)\cdot \left(1-\frac{a}{N_t}\right)\right\}.
\label{eq:expectedpopsize}
\end{equation}
Thus, below the critical population size and above the carrying capacity, individuals produce on average less than one offspring, whereas at intermediate population sizes individuals produce on average more than one offspring and the population is expected to grow (figure \ref{fig:Alleemodel}). We compared populations with critical population size $a=a_{AE} > 0$ to those without Allee effect, i.e. with critical size $a=0$. 
In all our analyses, the growth parameter $r$ takes values between 0 and 2, that is in the range where the carrying capacity $k_1$ is a locally stable fixed point of the deterministic Ricker model and there are no stable oscillations or chaos \citep[][p. 29]{101Vries2006}.

\subsection{Offspring-number models}
Our goal was to construct a set of offspring-number models that all lead to the same expected population size in the next generation (equation \eqref{eq:expectedpopsize}) but which represent a range of values for the variability in population dynamics and the strength of genetic drift $c$ (table \ref{tab:distributions}). All our models have in common that individuals are diploid and biparental, and, for simplicity, hermaphroditic. The models differ in how pairs are formed, in whether individuals can participate in multiple pairs, in whether or not selfing is possible, and in the distribution of the number of offspring produced by a pair.

\begin{table}
\centering
\caption{Properties of the offspring-number models considered in this study. The values for $c$, the relative strength of genetic drift in equilibrium, are derived in \ref{sec:genealogydetails}.}
\begin{tabular}{|l|l|l|}
\hline
model & $\mathbf{Var}[N_{t+1}|N_t]$ & relative strength of genetic drift $c$ \\
\hline
binomial & $\mathbf{E}[N_{t+1}|N_t] \cdot \left(1-\frac{\mathbf{E}[N_{t+1}|N_t]}{4 \cdot \lfloor N_t / 2 \rfloor}\right)$ & $\frac{3}{4}$ \\
Poisson & $\mathbf{E}[N_{t+1}|N_t]$ & 1 \\
Poisson-Poisson & $3 \cdot \mathbf{E}[N_{t+1}|N_t]$ & 2 \\
Poisson-geometric & $5 \cdot \mathbf{E}[N_{t+1}|N_t]$ & 3 \\
\hline
\end{tabular}
\label{tab:distributions}
\end{table}

\paragraph{Poisson model}
This is the model underlying the results in \otherpaper~and we use it here as a basis of comparison. Given a current population size $N_t$,
\begin{equation}
N_{t+1} \sim \text{Poisson}\left(\mathbf{E}[N_{t+1}|N_t]\right),
\end{equation}
such that $\mathbf{Var}[N_{t+1}|N_t]=\mathbf{E}[N_{t+1}|N_t]$.

\paragraph{Poisson-Poisson model}
Under this model, first a Poisson-distributed number of pairs
\begin{equation}
P_{t+1} \sim \text{Poisson}\left(\frac{1}{2} \cdot \mathbf{E}[N_{t+1}|N_t]\right)
\end{equation}
are formed by drawing two individuals independently, uniformly, and with replacement from the members of the parent generation. That is, individuals can participate in multiple pairs and selfing is possible. Each pair then produces a Poisson-distributed number of offspring with mean 2. The offspring numbers of the $P_{t+1}$ pairs are stored in the vector of family sizes $\mathbf{F_{t+1}}=(f_1,f_2,\dots,f_{P_{t+1}})$. This vector is required to simulate the genealogies backward in time, unlike in the Poisson  model where we only needed to store the total population size in each generation.

To compute the variance of $N_{t+1}$ given $N_t$, we used the formula for the variance of the sum of a random number of independent and identically distributed random variables \citep[][p. 13]{3100Karlin1975}
\begin{equation}
\mathbf{Var}[N_{t+1}|N_t]=\mathbf{E}[X]^2 \cdot \mathbf{Var}[P_{t+1}|N_t] + \mathbf{E}[P_{t+1}|N_t] \cdot \mathbf{Var}[X],
\label{eq:randomvariance}
\end{equation}
where $\mathbf{E}[X]$ and $\mathbf{Var}[X]$ are mean and variance of the number of offspring produced by a single pair. The resulting variance (see table \ref{tab:distributions}) is larger than that under the Poisson model.

\paragraph{Poisson-geometric model}
This model is identical to the Poisson-Poisson model except that the number of offspring of a given pair is geometrically distributed with mean 2, rather than Poisson-distributed. Using equation \eqref{eq:randomvariance} again, we obtain an even larger variance than under the Poisson-Poisson model (table \ref{tab:distributions}).

\paragraph{Binomial model}
Here, individuals can participate in only one pair and selfing is not possible. First, the individuals from the parent generation $t$ form as many pairs as possible, i.e. $P_{t+1}=\lfloor N_t / 2 \rfloor$. Then, each pair produces a binomially distributed number of offspring with parameters $n=4$ and $p = \frac{\mathbf{E}[N_{t+1}|N_t]}{4 \cdot P_{t+1}}$, such that the population size in the offspring generation
\begin{equation}
N_{t+1} \sim \text{Binom}\!\left(n=4 \cdot P_{t+1},\; p  = \frac{\mathbf{E}[N_{t+1}|N_t]}{4 \cdot P_{t+1}}\right).
\end{equation}
The reason for our choice of $n$ was that it leads to a variance $np(1-p)$ that is smaller than the variance under the Poisson model (see table \ref{tab:distributions}). As was the case for the previous two models, also here we needed to store the vector of family sizes to be able to simulate the genealogies backward in time.

\subsection{Demographic simulations}
As we are only interested in populations that successfully overcome demographic stochasticity and the Allee effect, we discarded simulation runs in which the new population went extinct before reaching a certain target population size $z$. Here, we used $z=2\cdot a_{AE}$, i.e. twice the critical population size in populations with Allee effect. We generated 20,000 successful populations with and without Allee effect for each offspring-number model and for a range of founder population sizes between 0 and $z$. The population-size trajectories $N_0,N_1,\dots,N_{T_z}$, where $N_{T_z}$ is the first population size larger or equal to $z$, and the family-size vectors were stored for the subsequent backward-in-time simulation of genealogies. We also used the population-size trajectories to compute the average number of generations that the 20,000 replicate populations spent at each population size before reaching $z$. The complete simulation algorithm was implemented in C++ \citep{3991Stoustrup1997}, compiled using the g++ compiler \citep{3992gcc}, and uses the boost random number library \citep{3993boost.org2013}. We used R \citep{575Team2009} for the analysis of simulation results.

\subsection{Simulation of genealogies}
From each successful model population, we simulated ten independent single-locus genealogies, each for ten individuals sampled at both genome copies at the time when the population first reaches $z$. To construct the genealogies, we trace the ancestral lineages of the sampled individuals backward in time to their most recent common ancestor. For the Poisson model, we applied the simulation strategy of \otherpaper: Given the population-size trajectory $N_0,N_1,\dots,N_{T_z}$  we let all lineages at time $t+1$ draw an ancestor independently, with replacement, and uniformly over all $N_t$ individuals in the parent generation. For the other offspring-number models considered in this study, we use a modified simulation algorithm (see \ref{sec:genealogydetails} for details) that takes into account the family-size information stored during the demographic simulation stage. Both simulation algorithms account for the possibility of multiple and simultaneous mergers of lineages and other particularities of genealogies in small populations. All lineages that have not coalesced by generation 0 are transferred to the source population. As in \otherpaper, we simulated this part of the ancestral history by switching between two simulation modes: an exact and a more efficient approximative simulation mode (see \ref{sec:genealogydetails}). At the end of each simulation run, we stored the average time to the most recent common ancestor $\overline{G_2}$ for pairs of gene copies in the sample.

To visualise our results and compare them among the offspring-number models, we divided $\overline{G_2}$ by the average time to the most recent common ancestor for two lineages sampled from the source deme ($2k_0/c$). The quotient $\overline{G_2}/(2k_0/c)$ can be interpreted as the proportion of genetic diversity that the newly founded population has maintained relative to the source population. We also computed the per-cent change in expected diversity in populations with Allee effect (AE) compared to those without:
\begin{equation}
\left(\frac{\overline{G_2} \text{ with AE}}{\overline{G_2} \text{ without AE}}-1\right)\cdot 100.
\label{eq:relchange}
\end{equation}

\subsection{Effective-size rescaled Poisson model}
Given a population size $n$ in an offspring-number model with relative strength of genetic drift $c$ (see table \ref{tab:distributions}), we define the corresponding effective population size as $n_e(n)=n/c$. In this way, a population of size $n$ in the target offspring-number model experiences the same strength of genetic drift as a Poisson population of size $n_e$. To approximate the various offspring-number models by a rescaled Poisson model, we thus set the population size parameters of the Poisson model ($a$, $k_0$, $k_1$, $z$, and $N_0$) to the effective sizes corresponding to the parameters in the target model. For example, to obtain a Poisson model that corresponds to the Poisson-geometric model we divided all population size parameters by 3.  
In cases where the effective founder population size $n_e(N_0)$ was not an integer, we used the next-larger integer in a proportion $q=n_{e}(N_0) - \lfloor n_{e}(N_0) \rfloor$ of simulations and the next-smaller integer in in the remainder of simulations. For the target population size, we used the smallest integer larger or equal to the rescaled value. All other parameters were as in the original simulations.

\section{Results}
\begin{figure}
  \includegraphics{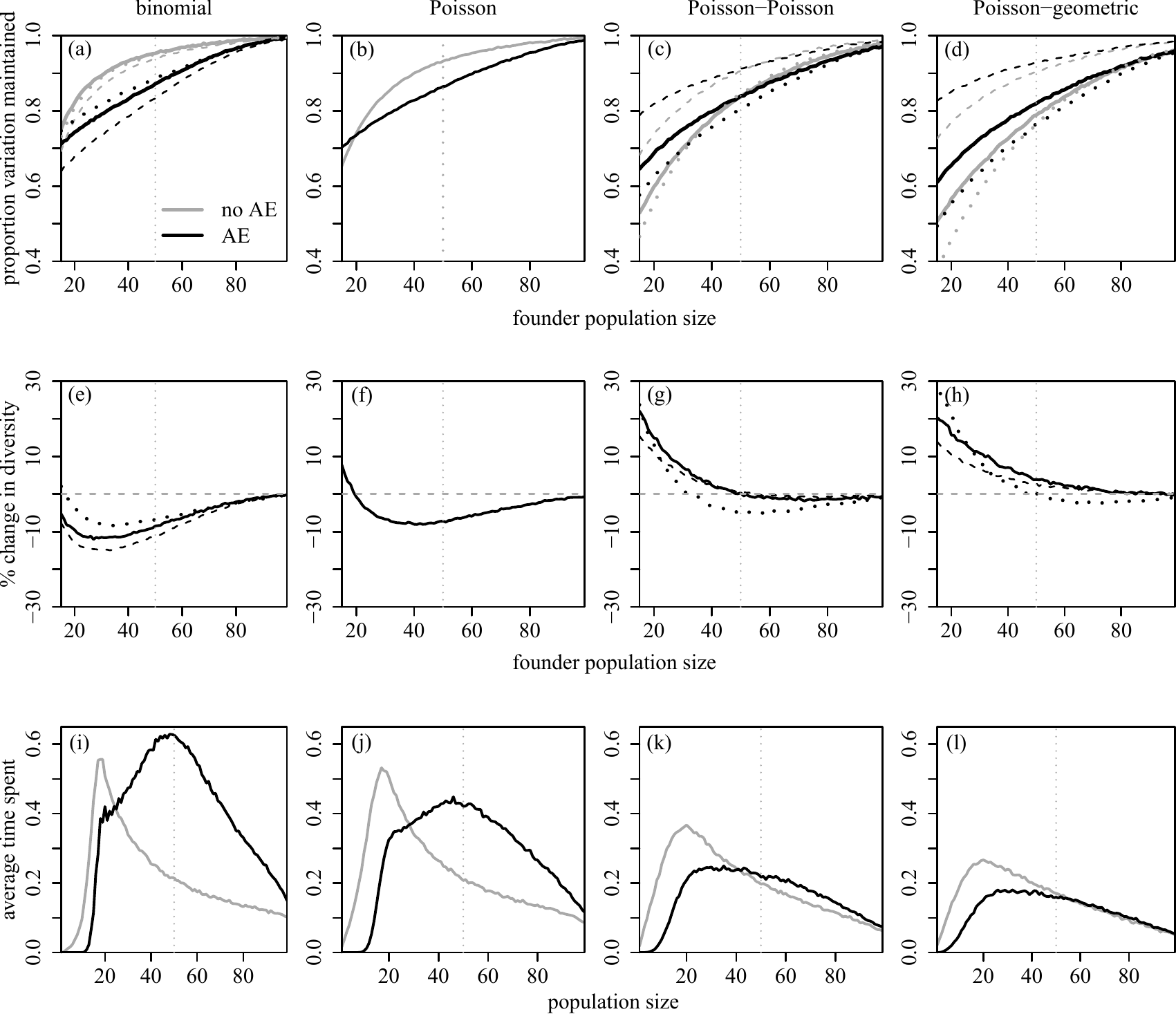}
  \caption{Consequences of the Allee effect for the population genetics and dynamics of successful populations under the binomial (1st column; a,e,i), Poisson (2nd column; b,f,j), Poisson-Poisson (3rd column; c,g,k) and Poisson-geometric model (4th column; d,h,l). Upper row: proportion of genetic variation maintained with Allee effect (AE, black lines) or without (no AE, grey lines). Middle row: per-cent change in genetic diversity in Allee-effect populations compared to those without Allee effect (see equation \eqref{eq:relchange}). Dashed lines in the upper two rows represent populations whose size trajectories were simulated from the respective offspring-number model, but where the genealogies were simulated assuming the Poisson model. Dotted lines show the results for the respective effective-size rescaled Poisson model. Lower row: average number of generations spent by successful populations at each of the population sizes from 1 to $z-1$ before reaching population size $z$, either with Allee effect (black lines) or without (grey lines). The founder population size for the plots in the lower row was 15. Note that in the case of the rescaled Poisson model, the values on the $x$-axis correspond to the founder population sizes before rescaling. Dotted vertical lines indicate the critical size of Allee-effect populations in the original model. Every point in (a--l) represents the average over 20,000 simulations. For the proportion of variation maintained, the maximum standard error of the mean was 0.0019. Parameters in the original model: $k_1=1000, k_0=10,000, z=100, r=0.1$.}
  \label{fig:results}
\end{figure}

The main results on the population dynamics and genetic diversity of populations with and without Allee effect are compiled in figure \ref{fig:results}. The upper two rows show the population genetic consequences of the Allee effect for different founder population sizes, the lower row the average number of generations that successful populations spend in different population-size ranges. Each column stands for one offspring-number model. Variation in offspring number and variability in the population dynamics increases from left to right. 
A first thing to note in figure \ref{fig:results} is that with increasing offspring-number variation the amount of genetic variation maintained in newly founded populations decreases, both for populations with and without Allee effect (solid black and grey lines in figure \ref{fig:results}a--d). In populations without Allee effect, however, the decrease is stronger. As a result, the magnitude and direction of the Allee effect's influence on genetic diversity changes as variation in offspring number increases. For the binomial model, the model with the smallest variability in population dynamics and genetics, the Allee effect has a negative influence on the amount of diversity maintained for all founder population sizes we considered (figure \ref{fig:results}a,e). For the model with the next-larger variation, the Poisson model, the Allee effect increases genetic diversity for small founder population sizes but decreases genetic diversity for large founder population sizes (figure \ref{fig:results}b,f). These results on the Poisson model are consistent with those in \otherpaper. As variability further increases, the range of founder population sizes where the Allee effect has a positive effect increases (figure \ref{fig:results}c,g). For the model with the largest offspring-number variation, the Poisson-geometric model, the Allee effect has a positive effect for all founder population sizes (figure \ref{fig:results}d,h). In summary, the larger is the offspring-number variation, the more beneficial is the Allee effect's influence on genetic diversity.

The differences between offspring-number models in the population genetic consequences of the Allee effect (represented by the solid lines in figure \ref{fig:results}a--h) result from two ways in which offspring-number variation influences genetic diversity: directly by influencing the strength of genetic drift for any given population-size trajectory, and indirectly by influencing the population dynamics of successful populations and thereby also the strength of genetic drift they experience. To disentangle the contribution of these two mechanisms, we first examine the direct genetic effect of offspring-number variation that results from its influence on the strength of genetic drift. For this, we generated a modified version for each of the binomial, Poisson-Poisson, or Poisson-geometric model (dashed lines in figure \ref{fig:results}). We first simulated the population dynamics forward in time from the original model. Backwards in time, however, we ignored this family-size information and let lineages draw their ancestors independently, uniformly, and with replacement from the parent generation as in the Poisson model. In the case of the binomial model, where the modified model has stronger genetic drift than the original model, both populations with and without Allee effect maintain on average less genetic variation in the modified than in the original model (figure \ref{fig:results}a). The Allee effect leads to a stronger reduction in genetic diversity in the modified model than in the original model (figure \ref{fig:results}e). The opposite pattern holds for the Poisson-Poisson and Poisson-geometric model where the modified model has weaker genetic drift than the original model. Populations in the modified model versions maintain a larger proportion of genetic variation (figure \ref{fig:results}c,d), and the relative positive influence of the Allee effect is weaker (figure \ref{fig:results}g,h).

Next, we consider the population dynamics of successful populations with and without Allee effect under the different offspring-number models. For this, we plotted the average number of generations that successful populations starting at population size 15 spend at each population size between 1 and $z-1$ before reaching the target state $z$ (lower row in figure \ref{fig:results}). As variability increases (going from left to right) both kinds of successful populations spend fewer generations in total, i.e. reach the target population size faster, but again populations with and without Allee effect respond differently to increasing variability. If we first focus on the offspring-number models with intermediate variation (figure \ref{fig:results}j,k) we observe that successful Allee-effect populations spend less time at small population sizes but more time at large population sizes than successful populations without Allee effect. This indicates that successful Allee-effect populations experience a speed-up in population growth at small sizes but are then slowed down at larger population sizes. If we now compare the results for the various offspring-number models, we observe that with increasing variability the speed-up effect becomes stronger and takes place over a larger range of population sizes, whereas the slow-down effect becomes weaker and finally disappears.

When rescaling the population-size parameters in the Poisson model to match one of the other offspring-number models, the resulting Poisson model behaves more similarly to the approximated model than does the original Poisson model, but the fit is not perfect (dotted lines in the upper two rows in figure \ref{fig:results}). In general, the model versions without Allee effect are better approximated by the rescaled Poisson models than the model versions with Allee effect. Although the proportion of variation maintained in the rescaled model is close in magnitude to the one in the target model, the rescaled model often differ in its predictions as to the genetic consequences of the Allee effect. Rescaled Poisson models always predict the Allee effect to have a positive effect for small founder population sizes and a negative effect for larger founder population sizes, although for the binomial and Poisson-geometric model the effect is always negative or positive, respectively (figure \ref{fig:results}e,h).

\section{Discussion}
\begin{figure}
 \centering
  \includegraphics{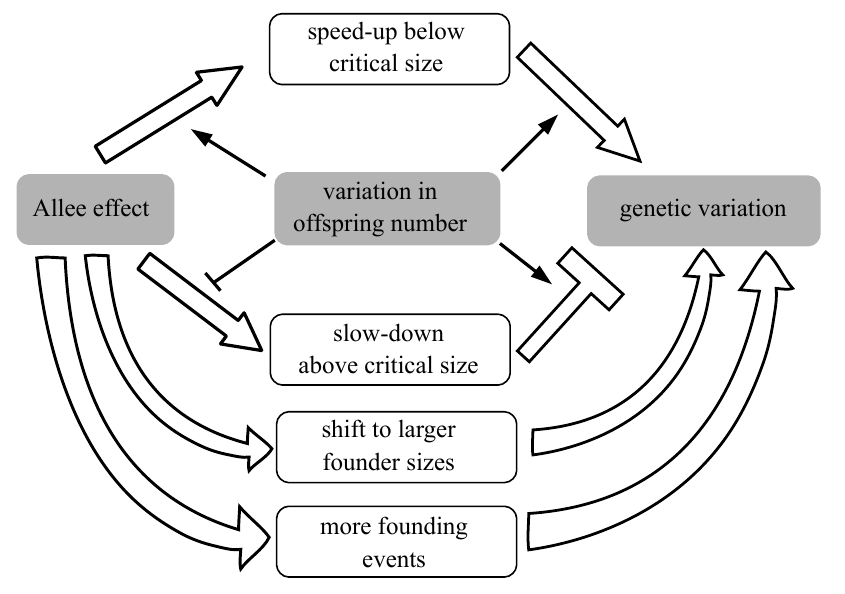}
  \caption{Overview over the various mechanisms by which the Allee effect influences the amount of genetic variation in successful introduced populations. Arrows represent positive effects while lines with bars represent inhibitory effects. Variation in offspring number enhances the speed-up of population growth caused by the Allee effect, but prevents the Allee-effect from causing a strong slow-down in population growth above the critical size. Variation in offspring number also magnifies the genetic consequences of both a speed-up and a slow-down in population growth.}
  \label{fig:flowchart}
\end{figure}

Our results indicate that offspring-number variation plays a key role for the genetic consequences of the Allee effect. We can understand a large part of the differences between offspring-number models if we consider how many generations successful populations spend in different population-size regions before reaching the target population size. In \otherpaper, we found that with Poisson-distributed offspring numbers successful Allee-effect populations spend less time at small population sizes than populations without Allee effect. Apparently, small Allee-effect populations can only avoid extinction by growing very quickly (speed-up in figure \ref{fig:flowchart}). We also found, however, that Allee effect-populations spend on average more time at large population sizes than populations without Allee effect (slow-down in figure \ref{fig:flowchart}). Consequently, under the Poisson model the Allee effect had either a positive or a negative effect on levels of genetic diversity depending on the founder population size. 

An increase in offspring-number variation leads to more variable population dynamics, which on the one hand lets successful populations escape even faster from the range of small population sizes than under the Poisson model. On the other hand, a large variation also prevents successful Allee-effect populations from spending much time near or above the critical population size because those that do still have a high risk of going extinct even at such high population sizes. Therefore, an increase in variability reinforces the speed-up effect but mitigates the slow-down effect and thus increases the range of founder population sizes for which the genetic consequences of the Allee effect are positive (see figure \ref{fig:results},\ref{fig:flowchart}). In that sense, variation in family sizes plays a similar role as variation in founder population size and in the number of introduction events (see figure \ref{fig:flowchart}), two factors that were examined in \otherpaper~ and, in the case of founder population size, also by Kramer and Sarnelle \citep{3699Kramer2008}. Variation in these aspects also leads to a positive influence of the Allee effect on diversity because by conditioning on success we let the Allee effect select the outliers of the respective distributions, and it is those outliers (particularly large founder sizes, exceptionally many introduction events) that lead to a large amount of genetic diversity.

Apart from its indirect but strong influence via the population dynamics, variation in offspring number also has a direct influence on genetic diversity by determining the strength of drift for a given population-size trajectory. Our comparisons between models with the same population dynamics but a different strength of drift suggest that an increase in the strength of genetic drift amplifies the per-cent change in diversity of Allee-effect populations compared to populations without Allee effect. We suggest this is the case because the stronger genetic drift is, the more genetic variation is lost or gained if it takes one generation more or less to reach the target population size. Thus, by reinforcing both the positive effect of a speed-up on genetic variation and the negative effect of a slow-down (figure \ref{fig:flowchart}), an increase in variation increases the magnitude of the net influence of the Allee effect on genetic variation.

We have now established that for a given set of parameter values, the population genetic consequences of the Allee effect differ strongly between offspring-number models. Nevertheless, we would still be able to use the Poisson model for all practical purposes if for any given set of parameters in one of the other offspring models we could find a set of effective parameters in the Poisson model that would yield similar results. The most obvious way to do this is to replace the population size parameters in the Poisson model by the corresponding effective population sizes, i.e. the population sizes in the Poisson model at which genetic drift is as strong as it is in the target model at the original population size parameter. However, our results (see figure \ref{fig:results}) indicate that the effective-size rescaled Poisson models cannot fully reproduce the results of the various offspring-number models. In particular, the population dynamics of successful populations remain qualitatively different from those under the target model. 

Apart from the population-size parameters, the Poisson model has an additional parameter that we could adjust, the growth parameter $r$. In \otherpaper, we considered different values of $r$ in the Poisson model. Small values of $r$ led to qualitatively similar results as we have seen here in models with a larger offspring-number variation and larger values of $r$. The reason appears to be that the population dynamics of successful populations depend not so much on the absolute magnitude of the average population growth rate (the deterministic forces) or of the associated variation (the stochastic forces), but on the relative magnitude of deterministic and stochastic forces. Indeed, a theorem from stochastic differential equations states that if we multiply the infinitesimal mean and the infinitesimal variance of a process by the same constant $\rho$, we get a process that behaves the same, but is sped up by a factor $\rho$ \citep[][Theorem 6.1 on p. 207]{3975Durrett1996}. Intuitively, we can make a model more deterministic either by increasing the growth parameter or by decreasing the variance. This suggests that we can qualitatively match the population dynamics of any given offspring-number model if we choose $r$ appropriately. Furthermore, we can match the strength of genetic drift if we rescale the population-size parameters appropriately. One could therefore suppose that by adjusting both the population-size parameters and the growth parameter in the Poisson model, we might be able to match both the population dynamic and the genetic aspects of the other offspring-number models. In \ref{sec:rescalingimpossible}, however, we show that this can only be possible if the equilibrium strength of genetic drift $c$ equals $\mathbf{Var}(N_{t+1}|N_t)/\mathbf{E}(N_{t+1}|N_t)$. This is not the case for the offspring-number models we examine in this study (see table \ref{tab:distributions}), neither with nor without Allee effect. However, our results suggest that the Allee effect enhances the mismatch between the effective-size rescaled Poisson models and their target models.

Overall, our results suggest that if we study populations that had been small initially but successfully overcame an Allee effect, microscopic properties such as the variation in offspring number can play a large role, although they may not influence the average unconditional population dynamics. Thus the common practice of first building a deterministic model and then adding some noise to make it stochastic may not produce meaningful results. As emphasised by Black and McKane \citep{3654Black2012}, stochastic population dynamic models should be constructed in a bottom-up way starting with modelling the relevant processes at the individual level and then deriving the resulting population dynamics in a second step. This means, we have to gather detailed biological knowledge about a species of interest before being able to predict the population genetic consequences of the Allee effect or other phenomena involving the stochastic dynamics of small populations. In this study, we have combined offspring-number models with a separate phenomenological model for the Allee effect. This helped us to compare the population-genetic consequences of the Allee effect among the different offspring-number model. A next step to make our model more mechanistic could be to let both the Allee effect and offspring-number variation emerge from a detailed mechanistic model for individual reproductive success.

In this study and in \otherpaper, we have focused on how the Allee effect impacts levels of neutral genetic diversity. Additional complications will arise if we take into account aspects of genetic variation that are required for the adaptation of an introduced population to its new environment. This need for adaptation can give rise to an Allee effect in itself because larger founding populations will have a larger probability to harbour the alleles that are advantageous in the new environment. We are currently investigating how this genetic Allee effect interacts with mate-finding or other ecological Allee effects and with the phenomena discussed in this paper.

\section{Acknowledgements}
We thank Peter Pfaffelhuber for pointing us to an interesting result on rescaling diffusion processes. MJW is grateful for a scholarship from the Studienstiftung des deutschen Volkes.

\setcounter{section}{0}
\renewcommand{\thesection}{Appendix \arabic{section}}
\renewcommand{\thefigure}{A\arabic{figure}}
\renewcommand{\thetable}{A\arabic{table}}
\renewcommand{\theequation}{A\arabic{equation}}
\setcounter{equation}{0}
\setcounter{figure}{0}

\section{Details of the genealogy simulation}
\label{sec:genealogydetails}
To simulate a genealogy, we start with the $n_s=10$ individuals sampled at generation $T_z$ and then trace their ancestry backward in time generation by generation until we arrive at the most recent common ancestor of all the genetic material in the sample. At time $t+1$, for example, we may have $n_a$ individuals that carry genetic material ancestral to the sample. To trace the ancestry back to generation $t$, we make use of the family-size vector $\mathbf{F}_{t+1}$ stored during the forward-in-time simulation stage. We first assign each of the $n_a$ individuals to a family by letting them successively pick one of the families in proportion to their sizes, i.e. the entries $f_i$ of the family-size vector. After a family has been picked, the corresponding entry in the family-size vector is reduced by 1 to avoid that the same individual is picked twice. 

After all $n_a$ individuals have been assigned, all families that have been chosen at least once draw parents in generation $t$. Here, the models differ. In the Poisson-Poisson and Poisson-geometric model, each family picks two parents independently, with replacement, and uniformly over all individuals in the parent generation. That is, parents can be picked by several families, and they can even be picked twice by the same family, corresponding to selfing. In the binomial model, on the other hand, parents are chosen without replacement. Each parent can only be picked by one family and selfing is not possible. From there on, everything works as described in \citep{3989Wittmann2013}: The two genome copies of each ancestral individual split and each of the two parents (or possibly the same individual in the case of selfing) receives one genome copy. A coalescent event happens if the same genome copy in a parent receives genetic material from several children.

At time $0$, i.e. at the time of the introduction, all remaining ancestors are transferred (backwards in time) to the source population, which is of large but finite size $k_0$ at all times. We assume that the mechanism of pair formation and the distribution of the number of offspring per pair are the same in the source population and in the newly founded population. Therefore, we still need to take into account family-size information to simulate the genealogy backward in time. Since we did not generate the required family-size vectors during the forward-in-time simulation stage, we now simulate them ad-hoc at every backward-in-time step. We do this by sampling from the distribution of family sizes (Poisson or geometric with mean 2, or binomial with $n=4, p=1/2$) until we get a total number of $k_0$ individuals (truncating the last family if required). Apart from this modification, the simulation algorithm is identical to the one used in the first stage of the simulation. When the simulation arrives at a point where all $n_a$ ancestors carry ancestral material at only one genome copy, it usually takes a long time until lineages are again combined into the same individual. Under these circumstances, we therefore use a more efficient approximative simulation mode: We draw the number of generations until two ancestors are merged into the same individual from a geometric distribution with parameter
\begin{equation}
p_{merge}=c \cdot \frac{\binom{n_{a}}{2}}{k_0},
\label{eq:coalapprox}
\end{equation}
where $c$ is the strength of genetic drift compared to that under the Poisson model (see table \ref{tab:distributions}). Whenever such an event happens, two randomly chosen ancestors are merged. Here we need to distinguish between two possible outcomes, both of of which occur with probability 1/2. In the first case, there is a coalescent event and we now have $n_a - 1$ ancestors, each still with ancestral material at only one genome copy. We therefore continue in the efficient simulation model. In the second case, there is no coalescent event and the resulting individual has ancestral material at both genome copies. Therefore, we switch back the other exact simulation mode. In this way, we switch back and forth between the two simulation modes until the most recent common ancestor of all the sampled genetic material has been found. For each simulation run, we store the average pairwise time to the most recent common ancestor of pairs of gene copies in the sample.

We will now take a closer look at the genealogical implications of the different offspring number models. Thereby, we will derive the values for the relative strength of genetic drift $c$ given in table \ref{tab:distributions} and  justify the approximation in \eqref{eq:coalapprox}. Equation \eqref{eq:coalapprox} is valid as long as we can neglect events in which at least three ancestors are merged into the same individual in a single generation (multiple mergers) and events in which several pairs of ancestors are merged in a single generation (simultaneous mergers). In the following, we will show that we can indeed neglect such events for large source population sizes $k_0$ because the per-generation probability of multiple and simultaneous mergers is $\mathcal{O}(1/k_0^2)$ whereas single mergers occur with probability $\mathcal{O}(1/k_0)$. Here $f(k_0)=\mathcal{O}(g(k_0))$ means that there exist positive constants $M$ and $k^*$ such that 
\begin{equation}
|f(k_0)| \leq M \cdot |g(k_0)|
\end{equation}
for all $k_0 > k^*$ \citep[][p. 107]{4129Knuth1997}. In other words, as $k_0$ goes to infinity, an expression $\mathcal{O}(1/k_0^n)$ goes to zero at least as fast as $1/k_0^n$.

To derive the probabilities of multiple and simultaneous mergers, we first introduce some notation. Let $P$ be a random variable representing the number of pairs with at least one offspring individual, and $O_l$ the number of offspring of the $l$th of them, with $l \in \{1,\dots,P\}$. Because we assume a fixed size $k_0$ of the source population and because we truncate the last family if necessary to achieve this (see above), the $O_l$ are not exactly independent and identically distributed. However, in all cases we are considering, $k_0$ is very large relative to typical offspring numbers $O_l$. To facilitate our argument, we therefore approximate the $O_l$ by independent and identically distributed $Y_l \text{ with } l \in \{1,\dots,P\}$ taking values in $\mathbb{N}$. 

In the following computations, we will need the fact that the first three moments of the $Y_l$ ($\mathbf{E}[Y_l],\: \mathbf{E}[Y_l^2], \text{ and } \mathbf{E}[Y_l^3]$) are finite. These are the moments of the number of offspring per pair, conditioned on the fact that there is at least one offspring. Let $X$ denote a random variable with the original distribution of the number of offspring per pair, i.e. the one allowing also for families of size 0. Then the moments of the $Y_l$ can be computed from the moments of $X$ with the help of Bayes' formula:
\begin{align}
\mathbf{E}[Y_l^m] & =\sum_{k=1}^\infty k^m \cdot \text{Pr}(X=k | X \geq 1) = \sum_{k=1}^\infty k^m \cdot \frac{\text{Pr}(X \geq 1|X=k)\cdot \text{Pr}(X=k)}{Pr(X\geq 1)} \nonumber \\
 & = \sum_{k=1}^\infty k^m \cdot \frac{\text{Pr}(X=k)}{1-\text{Pr}(X=0)} = \frac{\mathbf{E}[X^m]}{1-\text{Pr}(X=0)}.
\end{align}
These computations lead to finite constants for the first three moments of the $Y_l$ in the binomial, Poisson-Poisson, and Poisson-geometric model.

Since we switch to the accurate simulation mode whenever two lineages are combined into the same individual, we can assume here that the lineages at time $t+1$ are in different individuals. Thus, the number of lineages is equal to the number of ancestors $n_{a}$ and is at most twice the sample size (in our case at most 20), but usually much lower since many coalescent events already happen in the newly founded population. Of course, each ancestor has two parents, but since ancestors here carry genetic material ancestral to the sample only at one genome copy, only one of their two parents is also an ancestor and the other one can be neglected.

\begin{figure}
\includegraphics[width=\textwidth]{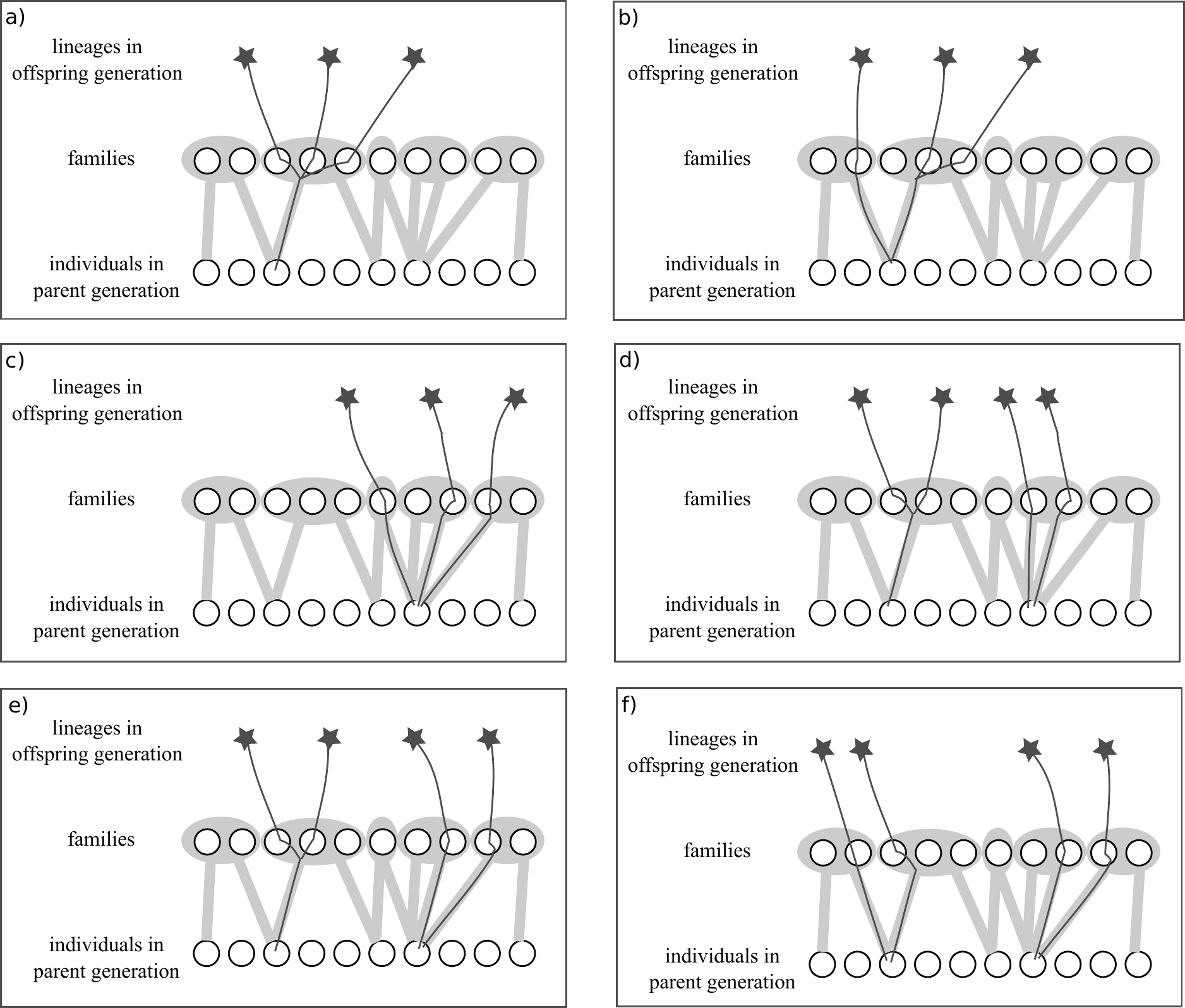}
\caption{Illustration of the various ways in which multiple and simultaneous mergers can arise. Here, the case of the Poisson-Poisson or Poisson-geometric model is depicted, where selfing is possible and individuals in the parent generation can contribute to multiple families.}
\label{fig:mergers}
\end{figure}

A multiple merger can occur in three, not necessarily mutually exclusive, ways (figure \ref{fig:mergers}a--c). First we will consider the case where there is at least one family that at least three lineages trace back to (figure \ref{fig:mergers}a). The probability of such an event is
\begin{align}
\text{Pr}\left(\max(Z_l) \geq 3\right) & \leq \mathbf{E}\left[\text{Pr}\left(\bigcup_{l=1}^P (Z_l \geq 3 ) {\bigg |} P\right) \right]  \nonumber \\
 & \leq \mathbf{E}\left[ \sum_{l=1}^P \text{Pr}(Z_l \geq 3) \right] \nonumber \\
 & \leq \mathbf{E}[P] \cdot \text{Pr}(Z_1 \geq 3) \nonumber \\
 & \leq k_0 \cdot \text{Pr}(Z_1 \geq 3)  \label{eq:triplemergerbound}\\
 & \leq k_0 \cdot \binom{n_{a}}{3} \cdot \left(\sum_{i=1}^\infty \text{Pr}(Y_1=i) \cdot \frac{Y_1}{k_0}\cdot \frac{Y_1-1}{k_0-1}\cdot \frac{Y_1-2}{k_0-2}\right) \nonumber \\
 & \leq k_0 \cdot \binom{n_{a}}{3} \cdot \frac{\mathbf{E}\left[Y_1^3\right]}{k_0^3}=\mathcal{O}\! \left(\frac{1}{k_0^2}\right).\nonumber
\end{align}
Here, we used the fact that $P \leq k_0$, which is true because we take $P$ to be the number of families with at least one member. Note that when lineages trace back to the same family, they still need to draw the same parent individual in order to be merged. For the three lineages here, this probability is 1/4 if the family did not arise from selfing, and 1 if it did. For simplicity, we take 1 as an upper bound for this probability here and for similar probabilities in the following inequalities.

Under the binomial model, the above case is the only way in which multiple mergers can occur. Under the Poisson-Poisson and Poisson-geometric model, however, parent individuals can participate in several pairs and therefore potentially contribute to more than one family. Therefore, we additionally have to take into account the possibility that lineages trace back to different families but then choose the same parent individual (figure \ref{fig:mergers}b,c). One possibility (figure \ref{fig:mergers}b) is that there is exactly one family that at least two lineages trace back to (event $E_1$), that two lineages in this family draw the same parent individual (event $E_2$), and that there is at least one lineage outside the family that draws the same parent (event $E_3$).
Using an argument analogous to that in \eqref{eq:triplemergerbound}, we obtain 
\begin{equation}
\text{Pr}(E_1) \leq \text{Pr}\left(\max(Z_l) \geq 2\right) \leq \binom{n_{a}}{2} \cdot \frac{\mathbf{E}[Y_1^2]}{k_0}.
\label{eq:E1}
\end{equation}
Furthermore, $\text{Pr}(E_2)\leq 1$ and
\begin{equation}
\text{Pr}(E_3) \leq (n_a-2) \cdot \frac{1}{k_0} \leq n_a \cdot \frac{1}{k_0}.
\end{equation}
Here we use the fact that families choose their parents independently and uniformly over all $k_0$ individuals in the parent generation, such that lineages in different families have a probability of $1/k_0$ of drawing the same parent. 
Combining the last three inequalities, we can conclude
\begin{equation}
\text{Pr}(E_1 \cap E_2 \cap E_3)  \leq \binom{n_{a}}{2} \cdot \frac{\mathbf{E}[Y_1^2]}{k_0^2} \cdot  n_{a} =  \mathcal{O}\!\left(\frac{1}{k_0^2}\right).
\end{equation}
Finally, the probability that lineages from at least three different families choose the same parent (figure \ref{fig:mergers}c) is bounded by
\begin{equation}
\binom{n_{a}}{3} \frac{1}{k_0^2}= \mathcal{O}\! \left(\frac{1}{k_0^2}\right).
\end{equation}

For a simultaneous merger to occur, there must be at least two mergers of at least two lineages each. This can happen in three (not necessarily mutually exclusive) ways (figure \ref{fig:mergers}d-f). To compute bounds for the corresponding probabilities, we again take 1 as an upper bound for all probabilities that lineages in the same family choose the same parent. First, a simultaneous merger can occur if there are at least two families, each with at least two lineages tracing back to them (event $M_1$, figure \ref{fig:mergers}d). This occurs with probability
\begin{align}
\text{Pr}(M_1) & \leq \mathbf{E}\Big[\text{Pr}\left( \exists \; l,m \in \{1,\dots,P\} \text{ s.t. } Z_l \geq 2 \text{ and } Z_m \geq 2 \big| P \right) \Big] \nonumber \\
 & \leq \mathbf{E}\left[ \sum_{l,m \leq P, \: l \neq m} \text{Pr}(Z_l \geq 2, Z_m \geq 2) \right] \nonumber \\
 & \leq \mathbf{E}\left[\binom{P}{2}\right] \cdot \text{Pr}(Z_1 \geq 2, Z_2 \geq 2) \\
 & \leq \binom{k_0}{2} \left(\sum_{i=1}^\infty \sum_{j=1}^\infty \text{Pr}(Y_1=i) \cdot \text{Pr}(Y_2=j) \cdot \binom{n_a}{2}^2 \cdot \frac{Y_1}{k_0} \cdot \frac{Y_1-1}{k_0-1} \cdot \frac{Y_2}{k_0-2} \cdot \frac{Y_2-1}{k_0-3}\right) \nonumber \\
 & \leq \frac{1}{2} \cdot \binom{n_a}{2}^2 \cdot \frac{\mathbf{E}[Y_1^2] \cdot \mathbf{E}[Y_2^2]}{(k_0-2)(k_0-3)} = \mathcal{O}\!\left(\frac{1}{k_0^2}\right)\nonumber .
\end{align}
Second, there can be a simultaneous merger if there is one family with at least two lineages tracing back to it and two lineages that merge separately without being in the family (event $M_2$, figure \ref{fig:mergers}e). Using the bound for the probability of exactly one merger of two individuals from \eqref{eq:E1}, we obtain
\begin{equation}
\text{Pr}(M_2)  \leq \text{Pr}(E_1) \cdot \binom{n_{a}}{2} \frac{1}{k_0} \leq \binom{n_{a}}{2}^2 \cdot \frac{\mathbf{E}\left[Y_1^2\right]}{k_0^2} = \mathcal{O}\!\left(\frac{1}{k_0^2}\right).
\end{equation}
Finally, we can have two pairs of lineages both merging without being in the same family (event $M_3$, figure \ref{fig:mergers}f). This occurs with probability
\begin{equation}
\text{Pr}(M_3) \leq \binom{n_{a}}{2}^2 \frac{1}{k_0^2}= \mathcal{O}\!\left(\frac{1}{k_0^2}\right).
\end{equation}

The different possibilities for multiple and simultaneous mergers are not mutually exclusive (for example there could be one triple merger and two double mergers), but since all of them have probability $\mathcal{O}\left(1/k_0^2\right)$, the probability of their union is also $\mathcal{O}\left(1/k_0^2\right)$. With these results, we can write
\begin{equation}
\text{Pr(1 merger)}=\binom{n_{a}}{2}\text{Pr}(A) + \mathcal{O}\!\left(\frac{1}{k_0^2}\right),
\label{eq:P1merger}
\end{equation}
where $A$ denotes the event that a specific pair of lineages merges into  the same individual in the previous generation.
\begin{align}
\text{Pr}(A) & = \text{Pr(same family)} \cdot \text{Pr}(A|\text{same family})\nonumber \\
 & + \Big(1-\text{Pr(same family)}\Big) \cdot \text{Pr}(A|\text{different families})
\label{eq:PrA}
\end{align}
For the binomial model,
\begin{equation}
\text{Pr}(A|\text{same family}) = \frac{1}{2}
\label{eq:Asamefambino}
\end{equation}
and
\begin{equation}
\text{Pr}(A|\text{different families})=0.
\label{eq:Adifffambino}
\end{equation}
For the other models,
\begin{equation}
\text{Pr}(A|\text{same family}) = \left(1-\frac{1}{k_0}\right) \cdot \frac{1}{2} + \frac{1}{k_0} \cdot 1 = \frac{1}{2}+ \frac{1}{2k_0},
\label{eq:Asamefamother}
\end{equation}
thereby accounting for the possibility of selfing, and
\begin{equation}
\text{Pr}(A|\text{different families})=\frac{1}{k_0}.
\label{eq:Adifffamother}
\end{equation}

For all models, we have
\begin{equation}
\text{Pr(same family)} = \frac{\mathbf{E}[S]}{k_0} = \frac{\mathbf{E}[X^*]-1}{k_0},
\label{eq:Psamefam}
\end{equation}
where $\mathbf{E}[S]$ is the expected number of siblings of a sampled individual and $\mathbf{E}[X^*]$ is the size-biased expectation of the number of offspring per family. Using \citep[][equation 4]{3975Arratia2010}, we obtain
\begin{equation}
\mathbf{E}[X^*]=\frac{\mathbf{Var}[X]}{\mathbf{E}[X]} + \mathbf{E}[X],
\end{equation}
which is 2.5 for the binomial model, 3 for the Poisson-Poisson model, and 5 for the Poisson-geometric model. 

Substituting \eqref{eq:Asamefambino}, \eqref{eq:Adifffambino}, and \eqref{eq:Psamefam} into  \eqref{eq:PrA} and then into \eqref{eq:P1merger}, we obtain for the binomial model:
\begin{equation}
\text{Pr(1 merger)} = \binom{n_a}{2} \cdot \frac{\mathbf{E}[X^*]-1}{2k_0} + \mathcal{O}\!\left(\frac{1}{k_0^2}\right).
\end{equation}
Analogously, substituting \eqref{eq:Asamefamother}, \eqref{eq:Adifffamother}, and \eqref{eq:Psamefam} into  \eqref{eq:PrA} and then into \eqref{eq:P1merger}, we obtain for the other models:
\begin{align}
\text{Pr(1 merger)} & = \binom{n_a}{2} \cdot \left[\frac{\mathbf{E}[X^*]-1}{k_0} \cdot \left(\frac{1}{2}+ \frac{1}{2k_0}\right) + \left(1-\frac{\mathbf{E}[X^*]-1}{k_0}\right) \cdot \frac{1}{k_0} \right] + \mathcal{O}\!\left(\frac{1}{k_0^2}\right)\nonumber \\
 & = \binom{n_a}{2} \cdot \left[\frac{\mathbf{E}[X^*]-1}{2k_0} + \frac{1}{k_0} \right]+ \mathcal{O}\!\left(\frac{1}{k_0^2}\right).
\end{align}
For the approximation in  \eqref{eq:coalapprox}, we neglect terms of order $1/k_0^2$ and thus the relative strength of genetic drift in \eqref{eq:coalapprox} is given by
\begin{equation}
c = \frac{\mathbf{E}[X^*]-1}{2} + \begin{cases} 0 & \text{for the binomial model} \\ 1 & \text{otherwise} \end{cases}.
\label{eq:cfinal}
\end{equation}
Evaluating this expression for the various models yields the values in table \ref{tab:distributions}.

\pagebreak

\section{Rescaling the Poisson model}
\label{sec:rescalingimpossible}

In this section, we argue that in general it is not possible to rescale the Poisson model such that it gives reasonable approximations to one of the other offspring-number models with respect to both population dynamics and genetics, even if we  both linearly rescale the population-size parameters and change the growth parameter. Given an offspring-number model $M$ with critical population size $a$, carrying capacity $k_1$, target population size $z$, founder population size $N_0$, growth parameter $r$, $\nu=\mathbf{Var}[N_{t+1}|N_t]/\mathbf{E}(N_{t+1}|N_t)$, and relative strength of genetic drift $c$, we will attempt to determine scaling parameters $s$ and $\rho$ such that a Poisson model $M'$ with parameters  $a'=s\cdot a,k_1'=s \cdot k_1,z'=s \cdot z,N_0'=s \cdot N_0$, and $r'=\rho \cdot r$ approximates the original model.

\begin{figure}
  \includegraphics[width=\textwidth]{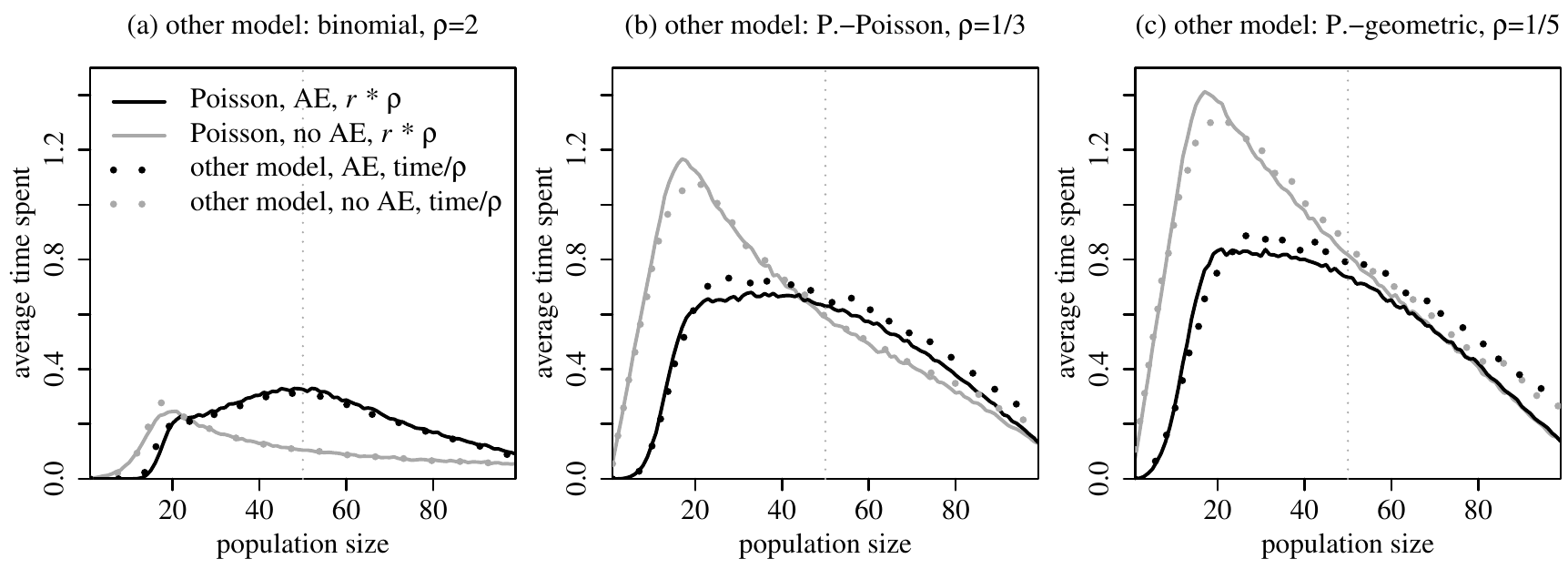}
  \caption{Comparison of Poisson models with growth parameter $r'=0.1 \cdot \rho$ to other offspring-number models, the binomial model (a), the Poisson-Poisson model (b), and the Poisson-geometric model (c), all with growth parameter $r''=0.1$. Each subplot shows the average number of generations that successful populations with Allee effect (AE) or without (no AE) under the various  models spend at different population sizes from 1 to $z-1$ before reaching the target population size $z$. In each case, we set $\rho=1/\nu$ (see table \ref{tab:distributions}, and using $\nu \approx 1/2$ for the binomial model) and divided the times spent at the different population sizes under the respective other offspring model by $\rho$ to account for the change in time scale. The dotted vertical line indicates the critical size $a$ of Allee-effect populations, here $a=50$. $k_1=1000, z=100$.}
  \label{fig:r_rescaling_timespent}
\end{figure}

In our argument, we will be guided by the following theorem on a time change in diffusion processes \citep[][Theorem 6.1 on p. 207]{3975Durrett1996}: If we speed up a diffusion process by a factor $\rho$, we obtain the same process as when we multiply both its infinitesimal mean and its infinitesimal variance by a factor $\rho$. In this study, we do not consider diffusion processes, but processes in discrete time and with a discrete state space. Furthermore, we cannot manipulate mean and variance independently. Therefore, the theorem cannot hold exactly for the models in this study. Nevertheless, it yields good approximations to the population dynamics under the the binomial, Poisson-Poisson, and Poisson-geometric model (figure \ref{fig:r_rescaling_timespent}). Specifically, a Poisson model $M'$ with  $r'=\rho \cdot r$ runs $\rho$~times as fast but otherwise exhibits approximately the same population dynamics as a model $M''$ with growth parameter $r$ and $\nu=\mathbf{Var}[N_{t+1}|N_t]/\mathbf{E}(N_{t+1}|N_t)=1/\rho$, given that all other parameters are the same, i.e. $a''=a',k_1''=k_1',z''=z'$ and $N_0''=N_0'$. However, due to the difference in time scale, the genetic drift experienced by populations under the two models may be very different.

To determine whether the offspring-number model $M$ can be approximated by a Poisson model $M'$, we will check whether it is possible to simultaneously fulfil two conditions, one on the population dynamics and one on the genetic aspect of the models. These two conditions are not sufficient to ensure that the models behave the same in every respect, but they appear necessary. If we can show that it is not possible to fulfil them simultaneously, not even in the unconditioned model, then the population dynamics and/or genetics of successful populations should be different under the two models.

First, we will specify a condition required to match the population dynamics. Since the success or failure of a population and other qualitative features of the population dynamics do not depend on the time scaling and since it is easier to compare models with the same growth parameter, we will use the model $M''$ instead of model $M'$ here. To match the relative strength of stochastic vs. deterministic forces in the population dynamics, we will require that the standard deviation of the population size in the next generation relative to the corresponding expected value is equal in both models for corresponding population sizes $n''=s \cdot n$:
\begin{equation}
\frac{\sqrt{\mathbf{Var}[N''_{t+1}|N''_t=n'']}}{\mathbf{E}[N''_{t+1}|N''_t=n'']}=\frac{\sqrt{\mathbf{Var}[N_{t+1}|N_t=n]}}{\mathbf{E}[N_{t+1}|N_t=n]}.
\label{eq:popdynamiccondition2}
\end{equation}
Given the properties of the model $M''$ and $M$ stated above, this is equivalent to
\begin{equation}
\frac{\sqrt{\frac{1}{\rho} \cdot \mathbf{E}[N''_{t+1}|N''_t=n'']}}{\mathbf{E}[N''_{t+1}|N''_t=n'']}=\frac{\sqrt{\nu \cdot \mathbf{E}[N_{t+1}|N_t=n]}}{\mathbf{E}[N_{t+1}|N_t=n]}.
\label{eq:popdynamiccondition3}
\end{equation}
\begin{equation}
\Leftrightarrow \frac{1}{\sqrt{\rho \cdot \mathbf{E}[N''_{t+1}|N''_t=n'']}}=\frac{\sqrt{\nu}}{\sqrt{\mathbf{E}[N_{t+1}|N_t=n]}}.
\label{eq:popdynamiccondition4}
\end{equation}
\begin{equation}
\Leftrightarrow \frac{1}{\sqrt{\rho \cdot n'' \cdot \phi\left(\frac{n''}{k_1''}\right)}}=\frac{\sqrt{\nu}}{\sqrt{n \cdot \phi\left(\frac{n}{k_1}\right)}},
\label{eq:popdynamiccondition5}
\end{equation}
where
\begin{equation}
\phi(x)=e^{r \cdot (1-x) \cdot \left(1-\frac{a}{k_1 \cdot x} \right)} = e^{r \cdot (1-x) \cdot \left(1-\frac{a''}{k_1'' \cdot x} \right)}
\end{equation}
is the expected per-capita number of surviving offspring in a population whose current size is a fraction $x$ of the carrying capacity (see equation \eqref{eq:expectedpopsize}). Since $n''/k_1'' = n/k_1$, \eqref{eq:popdynamiccondition5} reduces to 
\begin{equation}
\frac{1}{\nu}=\frac{\rho \cdot n''}{n} = \rho \cdot s.
\label{eq:popdynamiccondition6}
\end{equation}
This is our first condition.

Second, both models should have the same strength of genetic drift at corresponding population sizes $n$ and $n'$. Specifically, we require that the heterozygosity maintained over a corresponding time span is equal in both models:
\begin{equation}
\left(1-\frac{1}{n'}\right)^{1/\rho} = \left(1-\frac{c}{n}\right),
\end{equation}
which corresponds approximately to the condition
\begin{equation}
\frac{1}{\rho \cdot n'}=\frac{c}{n}
\label{eq:driftconditon}
\end{equation}
as long as $n$ and $n'$ are not too small. Here, we need the exponent $1/\rho$ because---as we have seen above and in figure \ref{fig:r_rescaling_timespent}---multiplying the growth parameter by a factor $\rho$ effectively speeds up the process by the same factor such that there is less time for genetic drift to act. Using $n'=s \cdot n$, \eqref{eq:driftconditon} simplifies to
\begin{equation}
\frac{1}{c} =\rho \cdot s. 
\label{eq:driftcondition2}
\end{equation}
This is our second condition.

Combining  \eqref{eq:popdynamiccondition6} and \eqref{eq:driftcondition2} shows that the two conditions can only be fulfilled simultaneously if $\nu = c$, which is not the case for the offspring-number models we consider in this study (see table \ref{tab:distributions}). The mismatch between $\nu$ and $c$ in our models is related to the way in which the diploid individuals form pairs to sexually reproduce. In a haploid and asexual model, in which individuals independently produce identically distributed numbers of offspring, $\mathbf{Var}[N_{t+1}|N_t]=N_t \cdot \mathbf{Var}[X]$ and $\mathbf{E}[N_{t+1}|N_t]=N_t \cdot \mathbf{E}[X]$, where $X$ is a random variable representing the number of offspring produced by a single individual. In analogy to \eqref{eq:cfinal}, we can then quantify the strength of genetic drift as
\begin{equation}
c = \mathbf{E}[X^*]-1 = \frac{\mathbf{Var}[X]}{\mathbf{E}[X]} + \mathbf{E}[X] -1 = \frac{\mathbf{Var}[N_{t+1}|N_t]}{\mathbf{E}[N_{t+1}|N_t]} + \mathbf{E}[X] -1 = \nu + \mathbf{E}[X] -1.
\end{equation}
This shows that in equilibrium, i.e. for $\mathbf{E}[X]=1$, there would be no mismatch between $c$ and $\rho$ in such a haploid model. In other situations, however, especially if we condition on the success of a small Allee-effect population, there could still be a mismatch. Furthermore, as discussed above, conditions  \eqref{eq:popdynamiccondition6}  and \eqref{eq:driftcondition2} may not be sufficient to ensure that two processes behave similarly. Especially if we condition on an unlikely event, the higher moments characterising the tail of the offspring-number distribution may be important and they are not necessarily matched even if mean and variance are. We therefore suggest that the strong differences among offspring-number distributions in the genetic consequences of the Allee effect can only in special cases be resolved by rescaling the parameters of the Poisson model.

\end{document}